\documentclass{XrU2005}
\usepackage{graphicx}

\title{The $XMM-Newton$ view of $\gamma-$ray loud active nuclei}

\author[1]{L. Foschini}
\author[2]{G. Ghisellini}
\author[3]{C.M. Raiteri}
\author[2]{F. Tavecchio}
\author[3]{M. Villata}
\author[1]{M. Dadina}
\author[1]{G. Di Cocco}
\author[1]{G. Malaguti}
\author[2]{L. Maraschi}
\author[4]{E. Pian}
\author[2]{G. Tagliaferri}

\affil[1]{INAF/IASF-Bologna, Via Gobetti 101, 40124 Bologna, Italy}
\affil[2]{INAF/Osservatorio Astronomico di Brera, Via Bianchi 46, 23807 Merate, Italy}
\affil[3]{INAF/Osservatorio Astronomico di Torino, Via Osservatorio 20, 10025 Pino Torinese, Italy}
\affil[4]{INAF/Osservatorio Astronomico di Trieste, Via G.B. Tiepolo 11, 34131 Trieste, Italy}

\begin{document}


\maketitle

\begin{abstract}
Notwithstanding the big efforts devoted to the investigation of the mechanisms responsible for the
high-energy ($E>100$~MeV) $\gamma-$ray emission in active galactic nuclei (AGN), the definite answer
is still missing. The X-ray energy band ($0.4-10$~keV) is crucial for this type of study, since both
synchrotron and inverse Compton emission can contribute to the formation of the continuum. Within an ongoing
project aimed at the investigation of the $\gamma-$ray emission mechanism acting in the AGN detected by
the EGRET telescope onboard CGRO, we firstly focused on the sources for which X-ray and optical/UV data 
are available in the \emph{XMM-Newton} public archive. The preliminary results are outlined here.
\end{abstract}

\section{Introduction}
The discovery of $\gamma-$ray loud AGN dates back to the dawn of $\gamma-$ray astronomy, when 
the European satellite \emph{COS-B} ($1975-1982$) detected photons in the $50-500$~MeV range from 
3C273 (Swanenburg et al. 1978). However, 3C273 remained the only AGN detected by \emph{COS-B}. 

A breakthrough in this research field came later with the Energetic Gamma Ray Experiment Telescope (EGRET) on board the 
\emph{Compton Gamma-Ray Observatory} (CGRO, 1991-2000). The third catalog of point sources contains $271$ sources 
detected at energies greater than $100$~MeV and $93$ of them are identified with blazars ($66$ at high confidence 
and $27$ at low confidence), and $1$ with the nearby radiogalaxy Centaurus~A (Hartman et al. 1999). Therefore, 
EGRET discovered that the blazar type AGN are the primary source of high-energy cosmic $\gamma-$rays (von Montigny 
et al. 1995). 

Later on, Ghisellini et al.~(1998) and Fossati et al.~(1998) proposed a unified scheme for $\gamma-$ray loud blazars,
based on their physical properties (see, however, Padovani et al. 2003). Specifically, the blazars are classified 
according to a sequence going from BL Lac to flat-spectrum radio quasar depending on the increase of the observed luminosity, 
which in turn leads to a decrease of the synchrotron and inverse Compton peak frequencies, and an increase of the ratio 
between the emitted radiation at low and high frequencies. In other words, the spectral energy distribution (SED)
of blazars is typically composed of two peaks, one due to synchrotron emission and the other to inverse Compton 
radiation. Low luminosity blazars have the synchrotron peak in the UV-soft X-ray energy band and therefore are ``high-energy
peaked'' (HBL). As the synchrotron peak shifts to low energies (near infrared, ``low-energy peaked'', LBL), 
the luminosity increases and the X-ray emission can be due to synchrotron or inverse Compton or a mixture of both.
For the Flat-Spectrum Radio-Quasars (FSRQ), the blazars with the highest luminosity, the synchrotron peak is in
the far infrared and the X-ray emission is due to inverse Compton. 

Moreover, the two-peaks SED is a dynamic picture of the blazar behaviour: indeed, these AGN are characterized by 
strong flares during which the SED can change dramatically. The X-ray energy band can therefore be crucial to 
understand the blazars behaviour and to improve the knowledge of high-energy emission. 

\section{Sample selection and data analysis}
To investigate the X-ray and optical/UV characteristics of $\gamma-$ray loud AGN in order to search for specific 
issues conducive to the $\gamma-$ray loudness, we cross correlated the $3^{\rm rd}$ EGRET Catalog (Hartman et al. 1999), 
updated with the identifications performed to date, with the public observations available in the \emph{XMM-Newton} 
Science Archive to search for spatial coincidences within $10'$ of the boresight of the EPIC camera. Fourteen AGN have 
been found (Table~1) as of April $14^{\rm th}$, 2005, for a total of $43$ observations. For three of them there are 
several observations available: 15 for 3C~$273$, 6 for Mkn~$421$, 9 for PKS~$2155-304$. The data from $6$ sources 
of the present sample are analyzed here for the first time and, among them, one has never been observed in X-rays before 
(PKS~$1406-706$). 

Data from the EPIC camera (MOS, Turner et al. 2001; PN, Str\"uder et al. 2001) and the Optical Monitor (Mason et al. 2001) 
have been analyzed with \texttt{XMM SAS 6.1} and \texttt{HEASoft 6.0}, together with the latest calibration files available 
at April $14^{\rm th}$, 2005, and by following the standard procedures described in Snowden et al. (2004). In addition,
the Optical Monitor makes it possible to have optical/UV data simultaneous to X-ray for most of the selected sources,
with the only exception of PKS~$0521-365$, Mkn~$421$, and Cen~A. 

\begin{table}[!t]
\caption{Main characteristics of the observed AGN.}
\centering
\begin{tabular}{lllc}
\hline
3EG          & Counterpart     & Type$^{\mathrm{*}}$ & Redshift \\
\hline
J$0222+4253$ & $0219+428$      & LBL      & $0.444$   \\
J$0237+1635$ & AO~$0235+164$   & LBL      & $0.94$    \\
J$0530-3626$ & PKS~$0521-365$  & FSRQ     & $0.05534$ \\
J$0721+7120$ & S5~$0716+714$   & LBL      & $>0.3$    \\
J$0845+7049$ & S5~$0836+710$   & FSRQ     & $2.172$   \\
J$1104+3809$ & Mkn~$421$       & HBL      & $0.03002$ \\
J$1134-1530$ & PKS~$1127-145$  & FSRQ     & $1.184$   \\
J$1222+2841$ & ON~$231$        & LBL      & $0.102$   \\
J$1229+0210$ & 3C~$273$        & FSRQ     & $0.15834$ \\
J$1324-4314$ & Cen A           & RG       & $0.00182^{\mathrm{**}}$ \\
J$1339-1419$ & PKS~$1334-127$  & FSRQ     & $0.539$  \\
J$1409-0745$ & PKS~$1406-076$  & FSRQ     & $1.494$  \\
J$1621+8203$ & NGC~$6251$      & RG       & $0.0247$ \\
J$2158-3023$ & PKS~$2155-304$  & HBL      & $0.116$  \\
\hline
\end{tabular}
\begin{list}{}{}
\item[$^{\mathrm{*}}$] \footnotesize{LBL: low frequency peaked BL Lacertae Object;
HBL: high frequency peaked BL Lacertae Object; FSRQ: flat-spectrum radio quasar; RG: radio galaxy.}
\item[$^{\mathrm{**}}$] \footnotesize{This redshift is not indicative and the distance of $3.84$~Mpc is adopted here.}
\end{list}
\label{tab:host}
\end{table}

\section{Main Results}
The main findings of this study can be summarized as follows:

(i) the EGRET blazars studied here have spectral characteristics in agreement with the unified sequence of 
Ghisellini et al. (1998) and Fossati et al. (1998);

(ii) no evident characteristics conducive to the $\gamma-$ray loudness have been found: the photon
indices are generally consistent with what is expected for this type of sources, with FSRQ that are harder 
than BL Lac; there are hints of some differences in the photon indices when compared with other larger catalogs (e.g.
\emph{BeppoSAX} Giommi et al. 2002), particularly for FSRQ: the sources best fit with a simple power law model 
show a harder photon index ($1.39\pm 0.09$ vs $1.59\pm 0.05$); however, the statistics is too poor to make firm 
conclusions (3 sources vs 26 in the \emph{BeppoSAX} catalog);

(iii) three sources show Damped Lyman $\alpha$ systems along the line of sight (AO~$0235+164$, PKS~$1127-145$,
S5~$0836+710$), but it is not clear if the intervening galaxies can generate gravitational effects altering 
the characteristics of the blazars so to enhance the $\gamma-$ray loudness;

(iv) no evidence of peculiar X-ray spectral features has been found, except for the emission lines of the 
iron complex in Cen~A.

More details of the analysis will be available in Foschini et al.~(2005).

\section*{Acknowledgments}
This work is based on public observations obtained with \emph{XMM--Newton}, an ESA science mission with instruments and
contributions directly funded by ESA Member States and the USA (NASA). This work was partly supported by the European 
Community's Human Potential Programme under contract HPRN-CT-2002-00321 and by the Italian Space Agency (ASI).

\end{document}